\documentclass[fleqn,usenatbib]{mnras}

\usepackage{newtxtext,newtxmath}
\usepackage{comment} 

\usepackage[T1]{fontenc}
\usepackage[normalem]{ulem}

\DeclareRobustCommand{\VAN}[3]{#2}
\let\VANthebibliography\thebibliography
\def\thebibliography{\DeclareRobustCommand{\VAN}[3]{##3}\VANthebibliography}

\usepackage{graphicx}	
\usepackage{amsmath}	

\title[Local Group isolation]{The Milky Way and M31 Orbital History: Did the Local Group evolve in isolation?} 

\author[Hartl \& Strigari]{
Odelia V. Hartl,$^{1}$\thanks{E-mail: odeliah@tamu.edu}
Louis E. Strigari,$^{1}$\thanks{E-mail: strigari@tamu.edu}
\\
$^{1}$Mitchell Institute for Fundamental Physics and Astronomy, Department of Physics and Astronomy, Texas A\&M University, College Station, TX, 77843 
\\
}

\date{Accepted 2025 January 29. Received 2025 January 28; in original form 2024 November 27}

\begin{document}
\label{firstpage}
\pagerange{\pageref{firstpage}--\pageref{lastpage}}
\maketitle

\begin{abstract}
We use new measurements of the M31 proper motion to examine the Milky Way (MW) - M31 orbit and angular momentum.  
For Local Group (LG) mass consistent with measured values, and assuming the system evolves in isolation, we show a wide range of orbits is possible. We compare to a sample of LG-like systems in the Illustris simulation, and find that $\sim 13\%$ of these pairs have undergone a pericentric passage. Using the simulated sample, we examine how accurately an isolated, two-body model describes the MW-M31 orbit, and show that $\sim 10\%$ of the analogues in the simulation are well-modeled by such an orbit. Systems that evolve in isolation by this definition are found to have a lower rate of major mergers, and in particular have no major mergers since $z \approx 0.3$. For all systems, we find an increase in the orbital angular momentum which is fairly independent of the merger rate, and is possibly explained by the influence of tidal torques on the LG. Given the likely quiet recent major merger history of the MW, it is plausible that the isolated two-body model appropriately describes the orbit, though recent evidence for a major merger in M31 may complicate this interpretation. 
\end{abstract}



\section{Introduction}

The Milky Way (MW) and M31 are the dominant constituents of the Local Group (LG). The sum of the masses of the MW and M31 dark matter halos appears to be consistent with the mass of LG as a whole~\citep{2021PhRvD.103b3009L,2022ApJ...928L...5B,Sawala:2022ayk}, though systematics such as the non-equilibrium nature of the MW and M31 may affect the interpretation of the mass estimators~\citep{2023ApJ...948..104P}; see~\citet{2020SCPMA..6309801W,2023arXiv230503293B} for recent reviews. A possible difference between the sum of the MW and M31 masses and the estimated LG mass as a whole may hint at additional, dark mass contained within the LG. 

An important, related property associated with the LG, MW, and M31 is the nature of its orbit. While it has long been assumed that the MW and M31 are on a radial approaching orbit~\citep{1959ApJ...130..705K,1974CoASP...6....7G,1981Obs...101..111L}, recent measurements of the 3D kinematics of M31 have challenged this standard assumption. In particular, proper motion measurements from Gaia DR2, EDR3 and the Hubble Space Telescope (HST) suggest a non-zero value for the MW-M31 relative tangential velocity~\citep{vanderMarel:2012xp,2019ApJ...872...24V,2021MNRAS.507.2592S,2024arXiv241018180R}. Though it still likely sub-dominant relative to the radial component, any tangential component of motion must be taken into account for mass estimates of the LG, and to understand the formation of the LG. 

In addition to these kinematic measurements, there has been a new understanding that galaxies within the LG, such as the Large Magellanic Cloud (LMC) or possibly even M33, have a significant effect on the kinematics of LG galaxies. In particular, the LMC may induce a relative motion between the MW disk and the extended MW stellar and dark matter halo, resulting in a displacement in the velocities between the disk and the halo components~\citep{2015ApJ...802..128G,2019ApJ...884...51G,2021Natur.592..534C,2021MNRAS.506.2677E}. ~\citet{2021NatAs...5..251P} use halo stars to measure this effect, and the resulting velocity shift adjusts the relative velocity between the MW and M31. This is in addition to the kinematic effect that changes the location of the MW-M31 barycenter due to a massive LMC~\citep{2016MNRAS.456L..54P}. 

Though it is a fundamental means to characterize the LG, and dictates the future fate of the MW and M31~\citep{2008MNRAS.386..461C,2024arXiv240800064S}, the above emphasizes how little is still known on the true nature of the MW-M31 orbit. The simplest model for the MW-M31 orbit appeals to two-body kinematics combined with cosmology. This model originally motivated~\cite{1959ApJ...130..705K} to examine the keplerian equation for a MW-M31 two-body orbit and use this to estimate the mass of the LG. The simple input assumptions were that the LG mass is dominated by the MW and M31, and that they can be modelled as spatially coincident near the time of the Big Bang. While the original analysis assumed a radial orbit, more generally the model may include tangential motion. Applying these assumptions along with the present day measured values for the relative separation and velocity, they reported a lower limit on the LG mass of $ \gtrsim 10^{12} \mathrm{M_{\odot}}$. This so-called timing model was calibrated to cosmological simulations which identify LG-like systems~\citep{1991ApJ...376....1K,2008MNRAS.384.1459L}. These analyses found that generally the timing mass estimator is reasonable, though biases may be incurred when accounting for a large tangential velocity and the cosmological constant~\citep{2014ApJ...793...91G,Hartl:2021aio}, and the aforementioned affect of the LMC~\citep{2022ApJ...928L...5B}. An appropriate definition of the halo mass may help to mitigate these biases~\citep{1991ApJ...376....1K,2023MNRAS.526L..77S}. 

While the assumption of a keplerian two-body orbit in which the LG evolves in isolation provides a reasonable starting point, it has yet to be  established that it accurately describes the LG. For example, major mergers, either within the MW or M31, may affect the LG kinematics. Though it appears the MW has had a relatively quiet recent major merger history, in which the last significant merger occurred $\gtrsim 8$ Gyr ago~~\citep{2018Natur.563...85H,2020MNRAS.494.3880B}, there is evidence for a more recent merger in M31~\citep{2018NatAs...2..737D,2018MNRAS.475.2754H}. The merger histories of LG-like systems may also be extracted from cosmological simulations~\citep{2020MNRAS.491.1531C}. Aside from mergers, tidal interactions between the LG and the local environment may influence the LG, for example via the influence of external torques that may alter the angular momentum. Though there has been progress in characterizing the matter distribution in the local volume of the universe~\citep{2012MNRAS.425.2049H}, its gravitational influence on the LG is still largely uncertain~\citep{2014MNRAS.440..405M}. 

In this paper, we undertake a critical examination of the MW-M31 orbit, in particular the assumption of isolation, the impact of mergers, and the angular momentum evolution of the orbit. Using the measured M31 kinematics and the estimates of the LG mass, we determine the possible orbits for the MW and M31 using the simple assumption of an isolated two-body orbit. Then, we determine how well this model describes the orbital history between the MW and M31 by comparing it to analogs in cosmological simulations, and we define a criteria for systems that are well-described by the two-body model. 

For the LG-analogue systems identified in the simulations, we determine the merger rates, and examine how this merger rate affects the orbital history of the LG. We additionally extract the angular momentum for our simulated sample, and compare to the observed orbital angular momentum of the LG. For the simulated sample, we examine the evolution of the angular momentum over cosmic time. 

This paper is organized as follows. In Section~\ref{sec:characteristics} we examine the possible orbits given the LG mass and kinematics. In Section~\ref{sec:simulations} we discuss the simulation sample used in our analysis, and the calculation of the merger rate in the simulations. In Section~\ref{sec:results} we present the results of our analysis, and then in the final section we move on to the conclusions and discussion. 

\section{Local Group Kinematic Properties and Orbital Characteristics}
\label{sec:characteristics} 

In this section, we review the kinematics properties of the LG, and use these to estimate the LG orbital angular momentum. We then move on to calculate possible LG orbits given the kinematic constraints. 

\subsection{LG mass and angular momentum}

The distance from the Sun to M31 is $770 \pm 40$ kpc, and the heliocentric velocity is $-301$ km/s~\citep{vanderMarel:2007yw}. For best fitting values of the MW circular and Local Standard Rest velocities~\citep{2011MNRAS.414.2446M}, this corresponds to a MW-centric radial velocity of $\mathrm{v_{rad}} = -110$ km/s. The results from Gaia EDR3 find that the tangential velocity is $\mathrm{v_{tan}} = 82.4 \pm 31.2  \mathrm{\, km \, s^{-1}}$ \citep{2021MNRAS.507.2592S}, which significantly rules out a purely radial orbit and is larger than previous measurements from HST~\citep{vanderMarel:2012xp}. Earlier results from Gaia DR2 are consistent with an even larger tangential velocity, $\mathrm{v_{tan}} = 170 \pm 51 \mathrm{\, km \, s^{-1}}$~\citep{2019ApJ...872...24V}. Systematics on these results have been considered in~\citet{2024arXiv241018180R}. Large tangential velocities may also be motivated by indirect measurements of the M31-MW relative motion~\citep{2014MNRAS.443.1688D,2016MNRAS.456.4432S}. 

The deduced radial and tangential velocities may be corrected to account for the impact of the LMC, which may induce a significant relative motion between the MW disk and the frame defined by the extended halo~\citep{2015ApJ...802..128G,2019ApJ...884...51G,2021Natur.592..534C,2021MNRAS.506.2677E}. Using simulations of LG-like systems,~\cite{2022ApJ...928L...5B} found that correcting for the LMC may increase the tangential velocity by $25-30 \mathrm{\, km \, s^{-1}}$,
increase the radial velocity by up to 50 $\mathrm{\, km \, s^{-1}}$, and increase the relative separation up to 40 kpc. This is consistent with the results from~\citet{2023ApJ...942...18C}, who use the measured ``travel velocity" of the MW disk with respect to the outer halo to find an upwards radial velocity correction of approximately 20 $\mathrm{\, km \, s^{-1}}$ and a corresponding increase in the tangential velocity by a few $\mathrm{\, km \, s^{-1}}$. 

We consider a range of mass estimates for the LG, motivated by the results from several independent measurements. Fitting to the measurements of the Local Hubble Flow,~\citet{2014MNRAS.443.2204P} obtain a LG mass of $(2.3 \pm 0.7) \times 10^{12}$ M$_\odot$ (which \citet{2017MNRAS.468.1300P} find is larger than the sum of the MW and M31 virial masses). Using a numerical least action method and fitting to orbits of LG galaxies,~\citet{2013ApJ...775..102P} obtain a LG mass of $(2 \pm 1) \times 10^{12}$ \, {\rm M}$_\odot$.~\cite{2014MNRAS.443.1688D} apply the virial theorem in combination with momentum conservation and obtain a LG mass of $(2.5 \pm 0.4) \times 10^{12} \, {\rm M}_\odot$. These results may be compared to a pure application of the timing argument combined with the measured tangential velocity from Gaia EDR3, which gives an estimated LG mass of $(5.8 \pm 1.0) \times 10^{12}$ M$_\odot$.These results are broadly consistent with measured values for the sum of the masses of the MW and M31~\citep{Sawala:2022ayk}, and may be also compared to results based on cosmological simulations~\citep{2014ApJ...793...91G,2020JCAP...09..056M,2021PhRvD.103b3009L,2022MNRAS.513.2385C}. Taking into account dynamical corrections due to the LMC, ~\citet{2022ApJ...928L...5B} find a total LG mass of $(3.7 \pm 0.5) \times 10^{12}$ M$_\odot$. Changing the embedding from $\lambda$ to $\ddot{a}/a$, \citet{2024A&A...689L...1B} found the lowest timing argument mass estimate of $~ 2 \times 10^{12}$ M$_{\odot}$. 

\label{sec:velocities}
\begin{figure}
    \centering 
	\includegraphics[width = 0.45\textwidth]{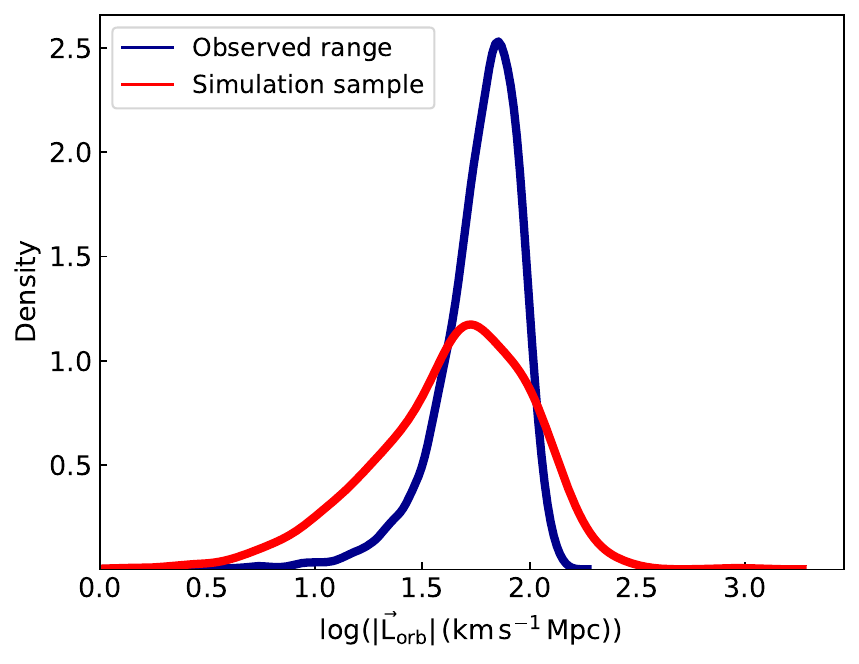}
   \caption{Density distribution of the orbital angular momentum calculated for all 597 pairs from the simulation sample at $z=0$, shown in the red curve. The blue curve shows the value and uncertainty in the angular momentum of the Local Group calculated from observed kinematics.}
    \label{fig:Lkdeonly}
\end{figure}

The orbital angular momentum per reduced mass for the MW/M31 system is $\vec L_{orb} = \vec r \times \vec{v}$, where $\vec r = \vec{r}_{MW} - \vec{r}_{M31}$ represents the separation vector, and $\vec v = \vec{v}_{MW} - \vec{v}_{M31}$ is the relative velocity. The magnitude of the orbital angular momentum is simply given by $|\vec L_{orb}| = r\mathrm{v_{tan}}$. Using the EDR3 measurement for the tangential velocity, the blue curve in Figure~\ref{fig:Lkdeonly} depicts the derived probability distribution for the magnitude of the orbital angular momentum. Here, the uncertainties on the proper motion are propagated to the orbital angular momentum, and the mean and the 90\% confidence interval of the distribution is given by 
\begin{equation} 
\mathrm{\log_{10}|\vec{L}_{orb} ( \, {\rm km} \, {\rm s}^{-1} \, {\rm Mpc} )| = 1.80_{-0.21}^{+0.14}}. 
\label{eq:Lorbdr3m31pm}
\end{equation}

\subsection{Analytic Orbits}
From the measured LG mass and angular momentum, we are in position to examine possible orbital trajectories. We consider a simplified analysis assuming that the MW and M31 evolve in isolation, and are unaffected by mergers or interactions with nearby structures. This serves as a simplified model which we can compare to the results from simulations in the following sections. 

\begin{figure*}
    \centering 
	\includegraphics[width = \textwidth]{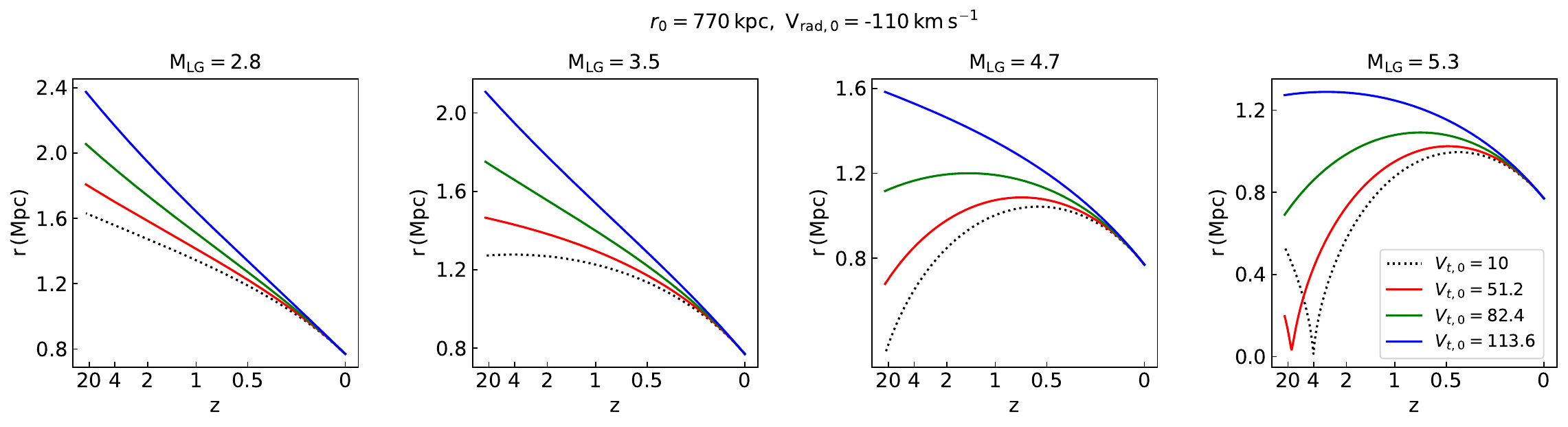}
 \includegraphics[width = \textwidth]{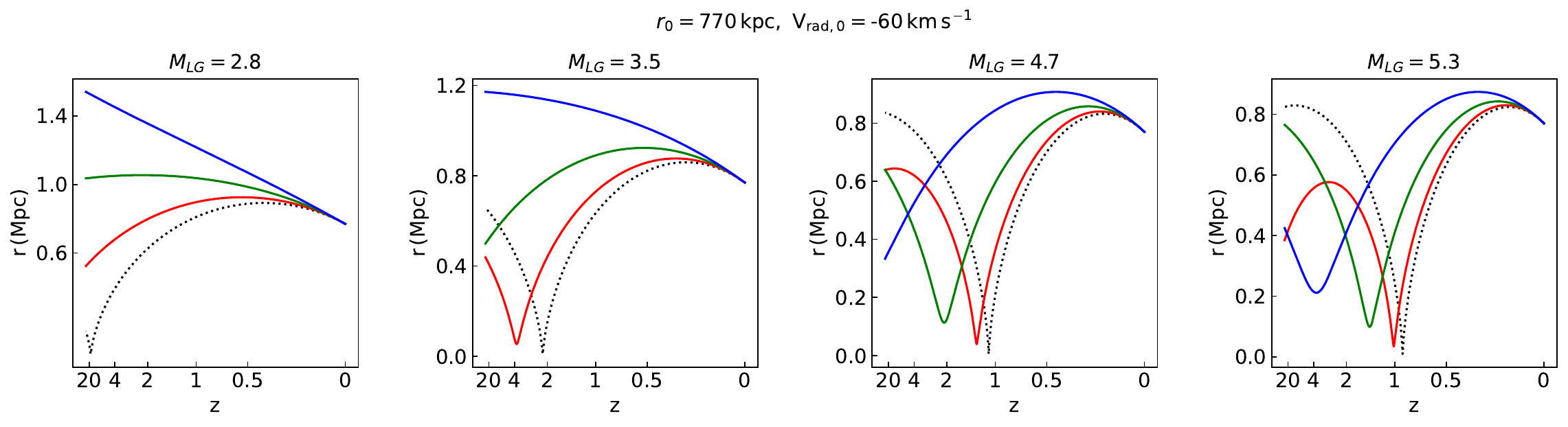}
    
    \caption{ 
    The MW/M31 orbital histories predicted using Equation~\ref{eq:MTAKEP} for different possibilities of present day values for the radial and tangential velocities as well as the Local Group mass. The tangential velocities are color-coded, as indicated in the legend, and the units are $\mathrm{km \, s^{-1}}$. The Local Group mass used in each case is denoted on the top x-axis in units of $10^{12} M_{\odot}$. In all plots the red, green, and blue curves show the orbit for different present day tangential velocities, ranging from the lower to higher estimated values, respectively. Additionally, both panels use the current day separation of $r_{0} = 770$ kpc.
    The top row shows the possible orbital histories predicted when using kinematic measurements that have not been adjusted for the possible correction due to the LMC, namely we utilize the present-day radial velocity of $\mathrm{v_{rad,0}} = -110 \, \mathrm{km \, s^{-1}}$. The bottom row utilizes the lower estimated radial velocity from the LMC correction. However, correcting for the LMC may also alter the relative separation between the MW and M31, increasing by about 40 kpc.}
    \label{fig:LG_orbits}
\end{figure*}

\begin{figure*}
\centering 
    \includegraphics[width = 0.49\textwidth]{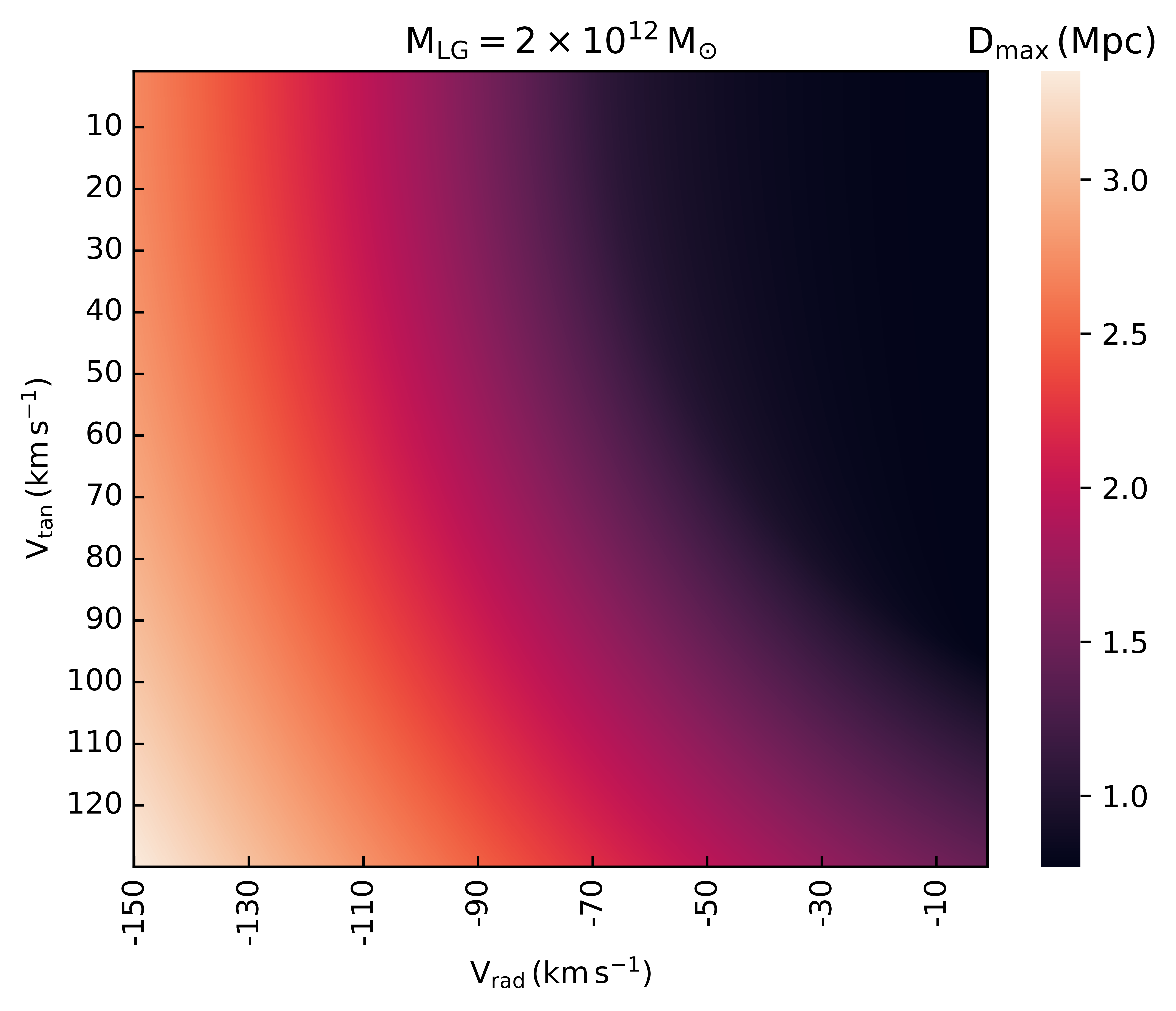}
    \includegraphics[width = 0.49\textwidth]{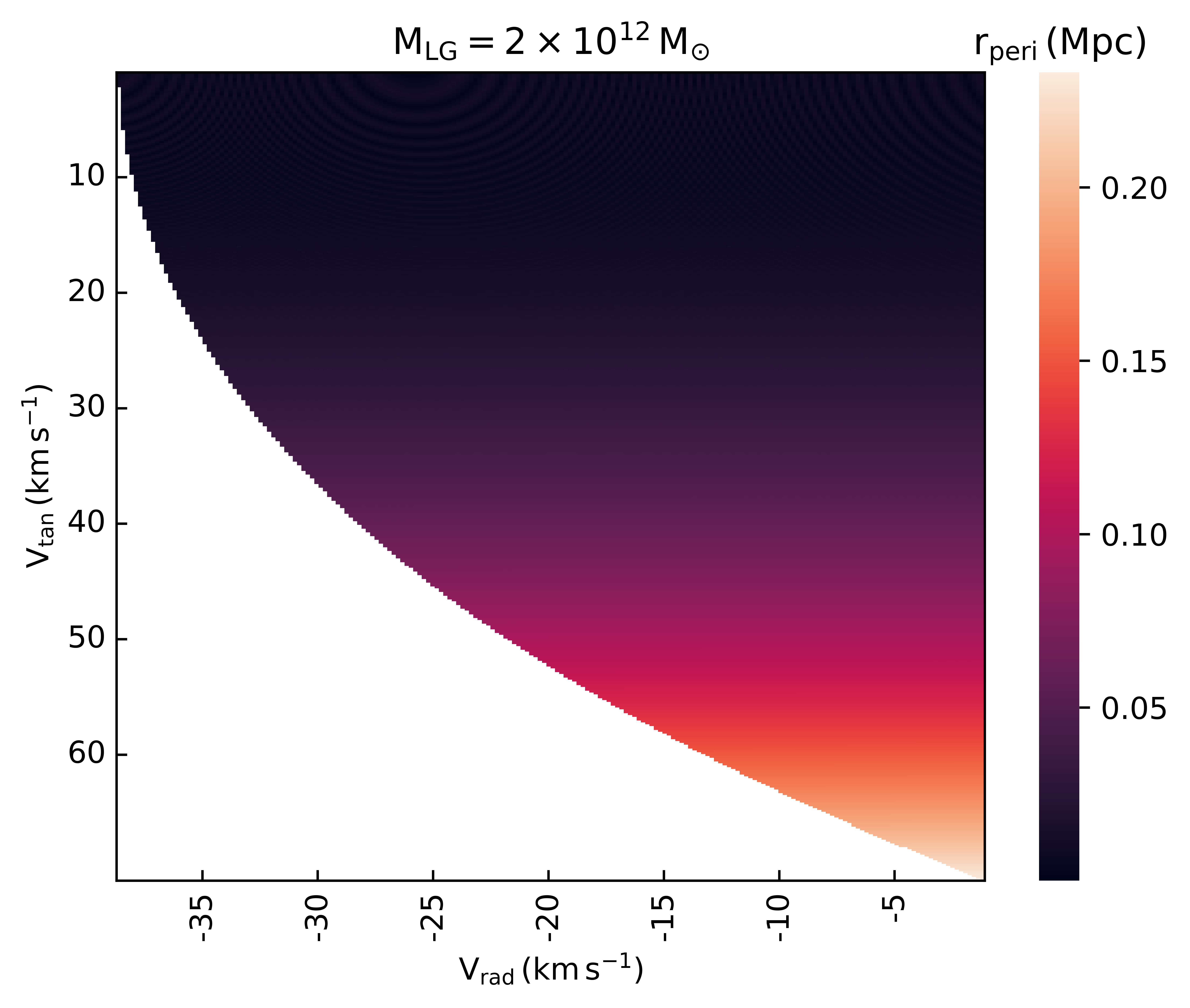}
\includegraphics[width = 0.49\textwidth]{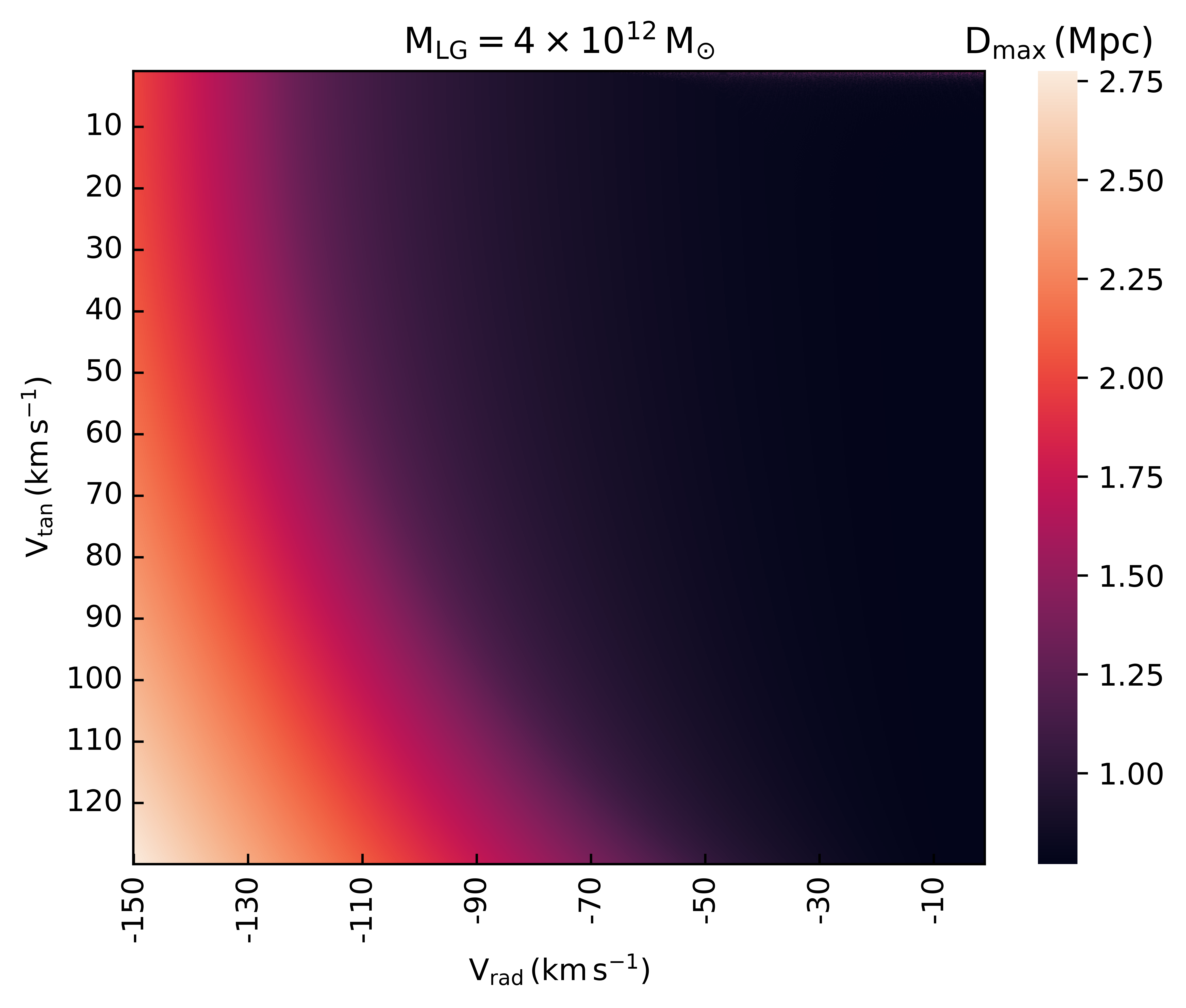}
\includegraphics[width = 0.49\textwidth]{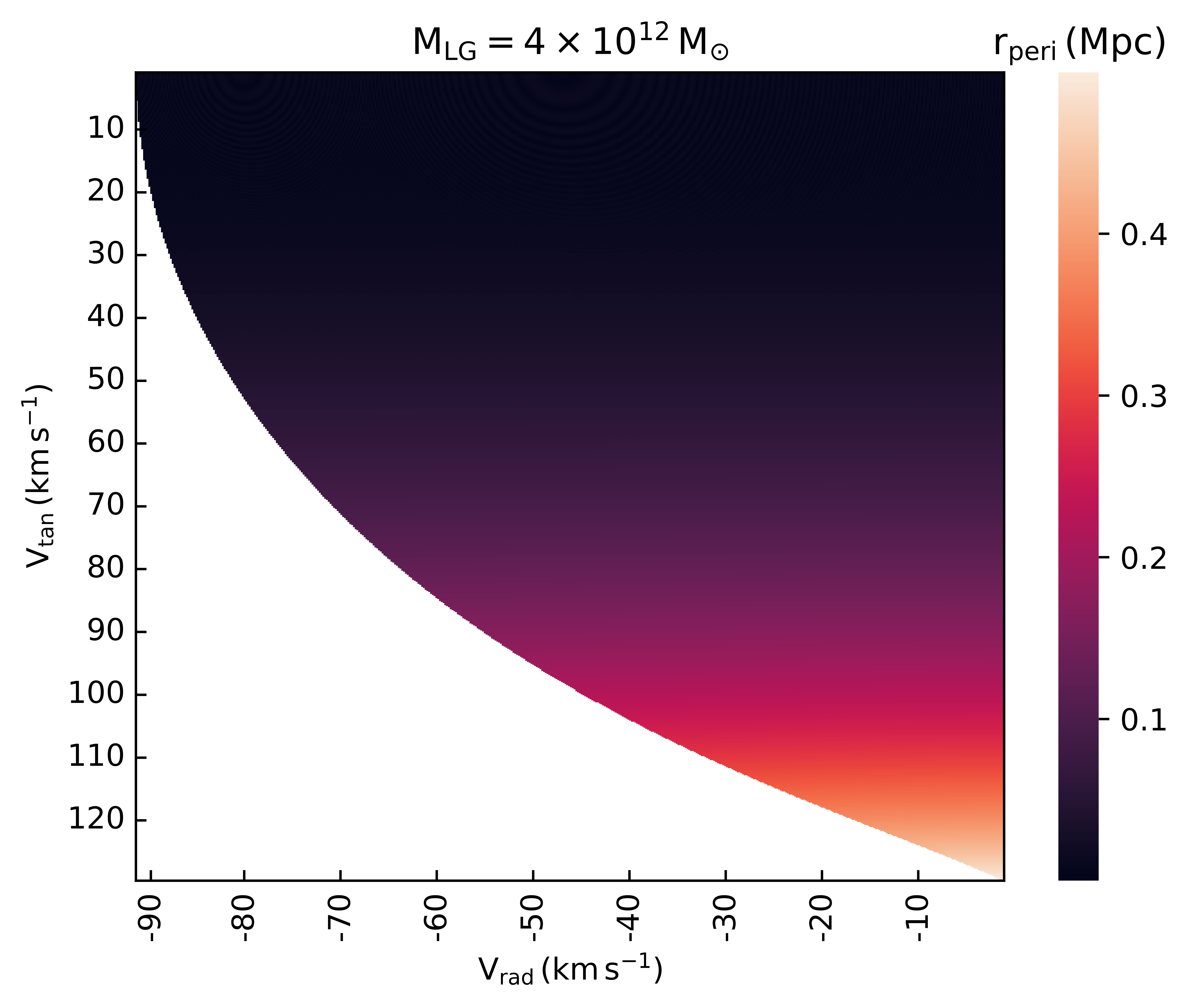}
    \caption{For a fixed present day Local Group mass of $2 \times 10^{12} M_{\odot}$ (top row) and $4 \times 10^{12} M_{\odot}$ (bottom row) and a present day separation of $770$ kpc, we solve the two-body equations for various combinations of present day radial and tangential velocity. All cases are plotted on the left, where the color bar represents the maximum separation between the pair. Pairs which are found to have undergone a pericentric passage are plotted on the right where the color bar denotes the distance of the pericentric passage. We see the distance of the pericentric passage is determined by the tangential velocity, and is independent of the radial velocity.  
    }
    \label{fig:M2_heatmap}
\end{figure*}

Defining the total mass as $M_{LG}=M_{MW}+ M_{M31}$, along with a polar coordinate system in which the separation $r$ defines the orbital plane, the equation of motion is 
\begin{equation} 
\ddot{r}= \frac{L^2}{r^3}-\frac{G M_{LG}}{r^{2}}+ H_0^2 \Omega_{\Lambda} r 
\label{eq:MTAKEP} 
\end{equation}
where $H_0 = 70$ km/s/Mpc is the Hubble constant, and $\Omega_{\Lambda} = 0.7$. Note that here we interpret $M_{LG}$ as the sum of the virial masses of the MW and M31, and as discussed below there is some ambiguity as to its precise definition. For example,~\citet{2023MNRAS.526L..77S} advocate for a definition of $M_{LG}$ as the mass contained within a sphere with radius defined by the present-day separation between the two galaxies. We use the range of measured angular momentum and LG mass and solve Equation~\ref{eq:MTAKEP} and plot resulting example two-body orbits in Figure~\ref{fig:LG_orbits}.  In all cases, we wind the orbit back to $z=20$, which matches the earliest snapshot of the simulation sample we compare to in Section \ref{sec:simulations}. The first row uses the canonical case in which the relative radial velocity is $\mathrm{v_{rad}} = -110 \mathrm{\, km \, s^{-1}}$. In this case, for the lower mass assumptions, the physical separation increases back to high redshift, implying that the galaxies are currently on their first approach. In this case, only for masses on the larger end of the allowed range do the two galaxies have a well-defined apocenter within the redshift range considered. For high mass and low tangential velocity, we see that a pericentric passage is even plausible at early times. 

The bottom row of Figure~\ref{fig:LG_orbits} show cases for which the radial velocity is increased (so the modulus of the velocity is decreased), which is motivated by the possible shift due to the LMC. Increasing the radial velocity to as large as $-60 \mathrm{\, km \, s^{-1}}$, plotted in the bottom row, for over half of the allowed mass range there is a pericentric passage for the range of the tangential velocities. Interestingly, even the median estimated mass shows a pericentric passage for the low tangential velocity system. This suggests that if the LMC significantly increases the relative radial velocity between the MW and M31, the pair is more likely to have undergone a pericentric passage. 

The above analysis is interesting when considered in the context of the classical timing model for the formation of the LG. This model relies on two assumptions: 1) $r \longrightarrow 0$ as $t \longrightarrow 0$, and 2) the LG is on its first infall. For the given set of present day kinematic measurements, the LG mass is then adjusted to satisfy the above boundary condition in the equation of motion. Focusing on the (red) set of curves in the first row of Figure \ref{fig:LG_orbits} with the lowest tangential velocity of $\sim 51$ km/s, which shows a pericentric passage in the rightmost panel, if we were to fit $M_{LG}$ to satisfy the timing mass conditions, we would recover a LG mass that is less than the true mass, as shown in the panel in the second column. The timing mass estimate would then systematically underestimate the true mass of the LG. The opposite extreme is illustrated for the case of the same tangential velocity and a lower LG mass, in which the two galaxies are on first infall, but had a larger separation at early times. If we fit the mass by enforcing the timing conditions, we would have estimated a larger LG mass, e.g. as given in the panel in the second column. The timing mass estimate would then overestimate the true mass of the LG. Note that this was the case found in~\citet{Hartl:2021aio} for the simulated sample discussed in the section below. Thus, we can conclude that it is likely the case that the two galaxies are on first infall and evolved from physical separations comparable to or larger than their present day separation. 

Regardless if the pair is on first infall or not, when we wind the orbits back we notice two populations of MW/M31 pairs: 1) those which evolved from separations smaller than their current separation, and 2) those which evolved from separations larger than their current separation. The first population reflects the canonical timing argument model, while the second population breaks with this traditional $r \longrightarrow 0$ as $t \longrightarrow 0$ assumption, in particular for systems with higher tangential velocity. Below we study this classification in more detail in the context of simulations. 

Figure \ref{fig:M2_heatmap} illustrates the general dynamics of two-body motion in terms of tangential and radial velocities. We consider systems with a current separation of $770$ kpc and masses of $2 \times 10^{12} M_{\odot}$ (top row) and $4 \times 10^{12} M_{\odot}$ (bottom row), and compute orbits for tangential velocities ranging from 0 to 130 $\mathrm{km \, s^{-1}}$ and radial velocities from -150 to 0 $\mathrm{km \, s^{-1}}$. The left plots are color-coded to display the maximum separation calculated for each scenario. The analysis indicates that the maximum separation is primarily influenced by the radial velocity. However, as the radial velocity decreases, the influence of tangential velocity on maximum separation becomes more pronounced. The right plots show a subset of the left plots, but focus on orbits that experience a pericentric passage, with color coding representing the distance of these passages. Notably, the distance of the pericentric passage is driven by the tangential velocity and appears independent of radial velocity, where a larger tangential velocity corresponds to a larger distance for these passages. A comparison between the top and bottom rows reveals that for a higher present-day LG mass, a broader range of radial and tangential velocities corresponds to systems that exhibit pericentric passages. 

To summarize the above analysis, due to the uncertainty in the LG kinematics, we deduce a wide range of possible orbits. As a next step, we will explore how well Equation~\ref{eq:MTAKEP}  predicts the orbits when known kinematic parameters are provided. We aim to answer these questions by analyzing a large sample of MW/M31 analog pairs in IllustrisTNG.

\section{LG-Analogs in simulations}
\label{sec:simulations} 
In this section, we describe the simulation sample used for our analysis, followed by a brief description of how their merger histories are calculated. 

\subsection{Simulation }
\par We identify a sample of LG-like systems in the Illustris TNG300-1 simulation \citep{2019ComAC...6....2N,2018MNRAS.475..624N, 2018MNRAS.475..648P, 2018MNRAS.475..676S, 2018MNRAS.477.1206N, 2018MNRAS.480.5113M}. The TNG300-1 simulation has a comoving volume of $302.6 \, {\rm Mpc}$ and $N=2500^{3}$ particles. This simulation includes baryonic physics with a particle mass of $m_{baryon}=7.6 \times 10^{6}$ M$_\odot h^{-1}$ and a dark matter particle mass of $m_{dm}=4.0 \times 10^{7}$ M$_\odot h^{-1}$. The simulation assumes cosmological parameters measured by Planck~\citep{2016A&A...594A...1P}.

The Illustris database stores galaxy properties at 100 snapshots. We utilize the position of each galaxy, defined as position of the particle with the lowest gravitational potential, and the velocity of each galaxy which is calculated by summing over all particles and weighting by the mass. The cuts imposed to obtain the sample of 597 LG-analogs, described in detail in~\citet{Hartl:2021aio}, were chosen to match the observed kinematics and stellar mass of the MW and M31. Briefly, we consider MW/M31 type galaxies as those with a B-band magnitude within $-22.3 < M_{B} < -19.3$, and identify approaching ($V_r < 0 \, \mathrm{km \, s^{-1}}$) MW-M31 analog pairs with a separation $500$ kpc $< |\Vec{r}| < 1000$ kpc, motivated by the  measured distance between the MW and M31. For the full motivation for the cuts used to obtain the sample, we refer to~\citet{Hartl:2021aio}. 

For the halo masses extracted from the simulations, we use the Subfind masses as provided in the Illustris database. This tends to be on average larger than that $M_{200}$ masses, defined as the mass within a sphere with a radius, $r_{200}$ that encompasses 200 times the critical density~\citep{2008MNRAS.384.1459L}. This is because the Subfind algorithm includes particles outside of $r_{200}$, though it excludes particles within subhalos of the main halo. 

\subsection{Mergers}
\label{sec:merger}

\begin{figure}
    \centering 
	\includegraphics[width = 0.50\textwidth]{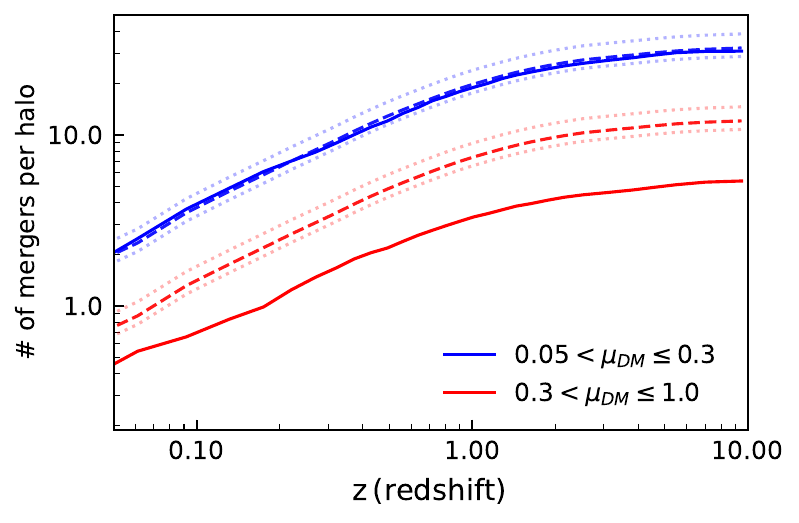}
   \caption{The cumulative merger history for all MW and M31 pairs, calculated as discussed in Section \ref{sec:merger}. The blue (upper) curves represent minor mergers while the red (lower) curves represent major mergers. The dashed and dotted curves represent the fitting function when applying the median and 90\% c.l of the mass distribution of the sample.
   }
    \label{fig:all_merger}
\end{figure}

We calculate the merger history of each galaxy in our sample with the merger catalogs constructed by the SubLink algorithm, see \citet{2015MNRAS.449...49R} for a detailed description of this process. Note, these authors use the merger trees following the stellar particles, and we use the merger trees following the dark matter particles. Additionally, they provide the details for calculating the merger rate. In determining the merger history, we classify each merger as either a major or a minor merger. The type of merger is determined by the merger mass ratio, which is specifically the ratio of the mass between the main halo and the halo which merged into the main halo. Major mergers are classified by a merger mass ratio $\mu$ of $0.3 < \mu \leq 1.0$, and minor mergers with $0.05 < \mu \leq 0.3$. Note that, as discussed in more detail below, the LMC impact into the MW is classified as a minor merger. 

Briefly, the merger rate,  $dN_m/dz$, is calculated in a few straightforward steps, where $N_{m}$ is the number of mergers and $dz$ is the redshift bin width. The steps are as follows: 
\begin{enumerate}
    \item Determine the range of merger mass ratios: $0.05<\mu_{DM} \leq 0.3$ (minor), $0.3<\mu_{DM} \leq 1.0$ (major).\\
    \item Define redshift bins.\\
    \item Count the number of mergers in each redshift bin.\\
    \item Divide by the redshift bin width. \\
    \item  Divide by the number of galaxies in the sample at $z=0  (1194)$. \\
\end{enumerate}

\citet{2010MNRAS.406.2267F} have explored the merger rate of all galaxies in the Millennium simulation for a halo of mass $M$, and reported a fitting function for the merger rate, given by  
\begin{equation}
\label{eq:Fakhouri_fit}
    \frac{dN_{m}}{d\mu_{x} dz} = A \left( \frac{M}{10^{12}M_{\odot}} \right)^{\alpha_{1}} \mu^{\beta_{1}} \exp{\left[ \left( \frac{\mu}{\bar{\mu}} \right) ^{\gamma} \right]}(1+z)^{\eta}. 
\end{equation}
The best fit parameters were found to be $(\alpha_{1} , \beta_{1}, \gamma ,\eta) = (0.133, -1.995, 0.263, 0.0993)$ and $(A, \bar{\mu}) = 0.0104, 9.72 \times 10^{-3}) $. Note that the halo mass, $M$, drives the behavior of this fitting function and depends on $z$. The redshift dependence of the halo mass is given by~\citep{2020MNRAS.491.1531C},
\begin{equation}
\label{M_z}
\frac{M_z}{M_{0}} = (1+z)^\beta \exp{[-\alpha(\sqrt{1+z} -1)]}
\end{equation}
with the best fit parameters of $\alpha$ and $\beta$ obtained for MW/M31 like galaxies. For our results, we use the M31(CS) value of $\alpha = 6.01$ and $\beta = 2.66$ from~\cite{2020MNRAS.491.1531C}. 

We compare the merger rate calculated from our sample of analog pairs to the prediction from the above fitting function. In performing this comparison, we can determine if MW/M31 type galaxies in pairs have a similar merger history to what is predicted for a typical MW/M31 mass galaxy.

In Figure \ref{fig:all_merger} we plot the cumulative form of the differential merger rate $dN_{m}/dz$, which represents the total number of mergers per halo from redshift $z=0$ out to redshift $z$. Figure \ref{fig:all_merger} shows that the majority of MW and M31-like galaxies in LG pairs have each had more than 10 minor mergers and a few major mergers since $z \sim 1$. For comparison, we plot the~\cite{2010MNRAS.406.2267F} fitting function using the median, 5\%, and 95\% points on the total mass distribution. Focusing on the minor merger history for MW/M31 in LGs, the blue curve in Figure~\ref{fig:all_merger} suggests our sample is in agreement with the~\citet{2010MNRAS.406.2267F} prediction. Note that, while the normalization of the major mergers is below the prediction, this is most likely explained by the scatter in the major merger rate between different simulations~\citep{2010ApJ...724..915H}. Importantly, the shape of the major merger is very similar to that of the~\citet{2010MNRAS.406.2267F} prediction.

\section{Results }
\label{sec:results}

We now move on to examine the orbital histories, angular momentum, and merger histories for our LG analogues. We determine how well the orbital histories match the simple, isolated two-body model discussed above, and examine how the properties of the analogues depend on their orbital histories. 

\subsection{MW-M31 orbital history in Illustris}

\begin{figure*}
\centering 
    \includegraphics[width = 0.49\textwidth]{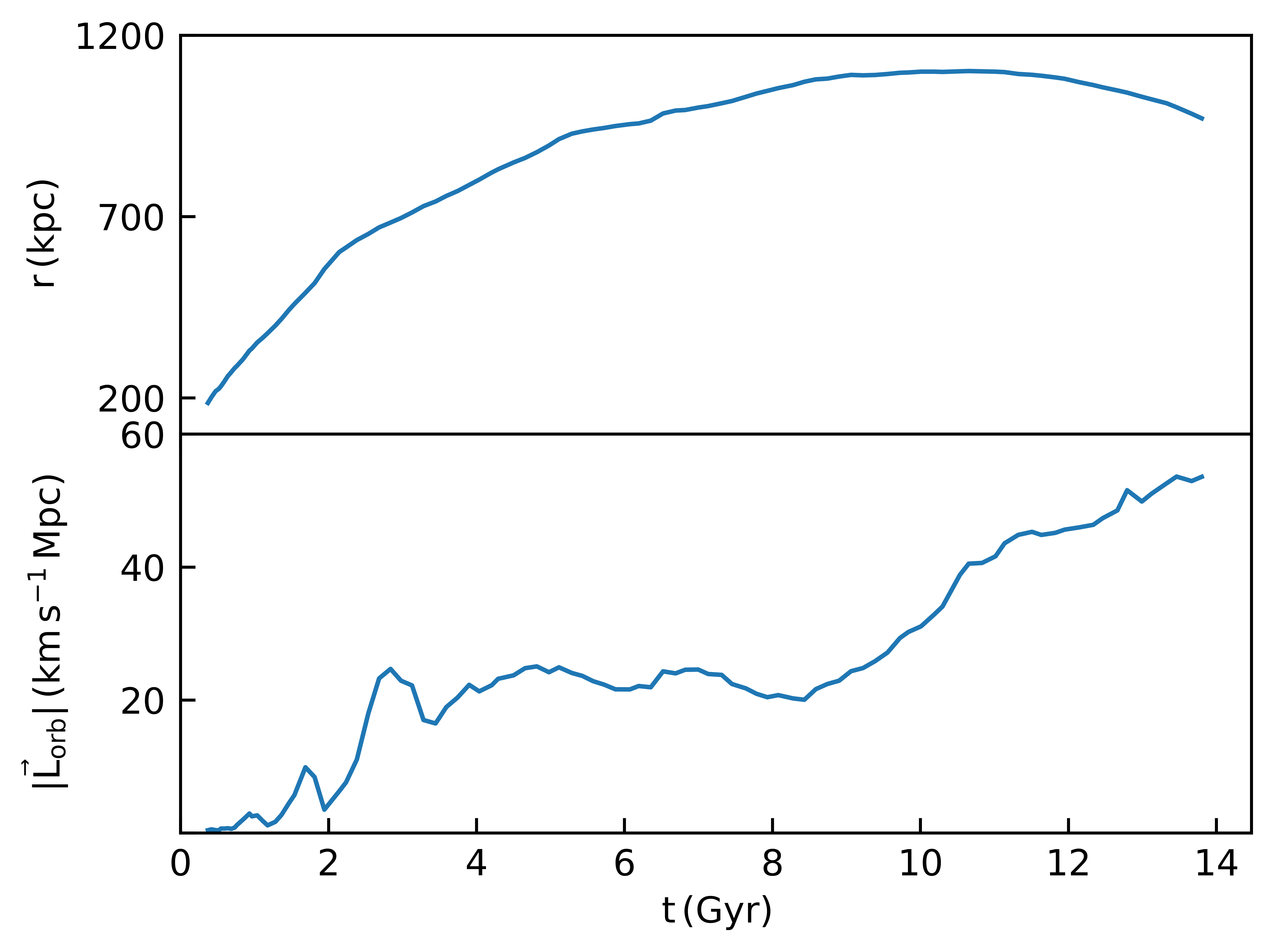}
    \includegraphics[width = 0.49\textwidth]{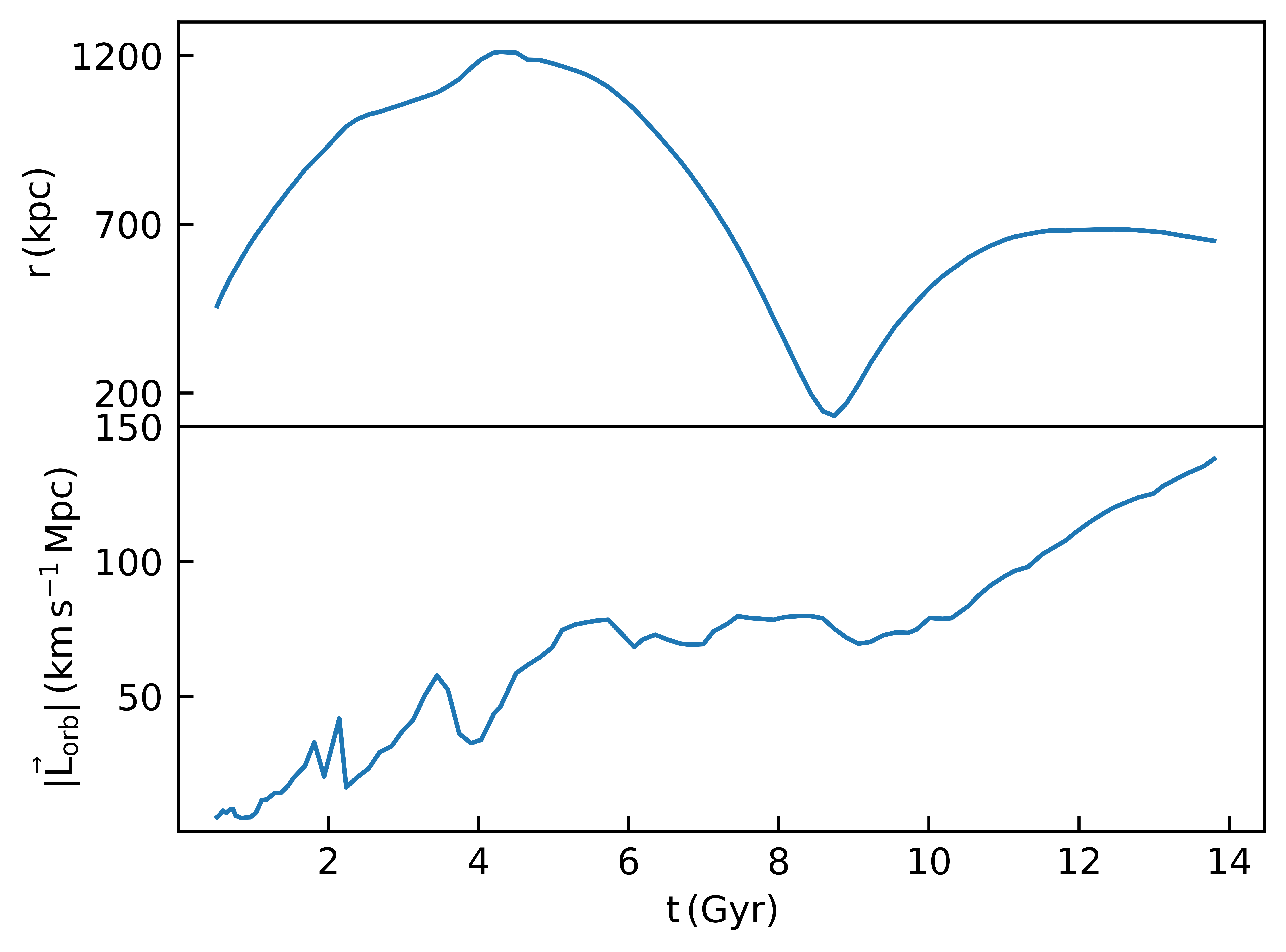}
    \caption{The orbital history for two LG-analogs in Illustris. The left plot shows an isolated two body infall, which perfectly reflects the timing argument model. The right plot shows a LG-analog which had a clear pericentric passage at $r = 131.7$ kpc when the universe was 8.7 Gyr old. In our sample we find 77 out of 597 LG-analogs have similarly undergone one pericentric passage. The remaining sample is on first infall. 
    }
    \label{fig:simorbits}
\end{figure*}

\begin{figure}
\centering 
    \includegraphics[width = 0.49\textwidth]{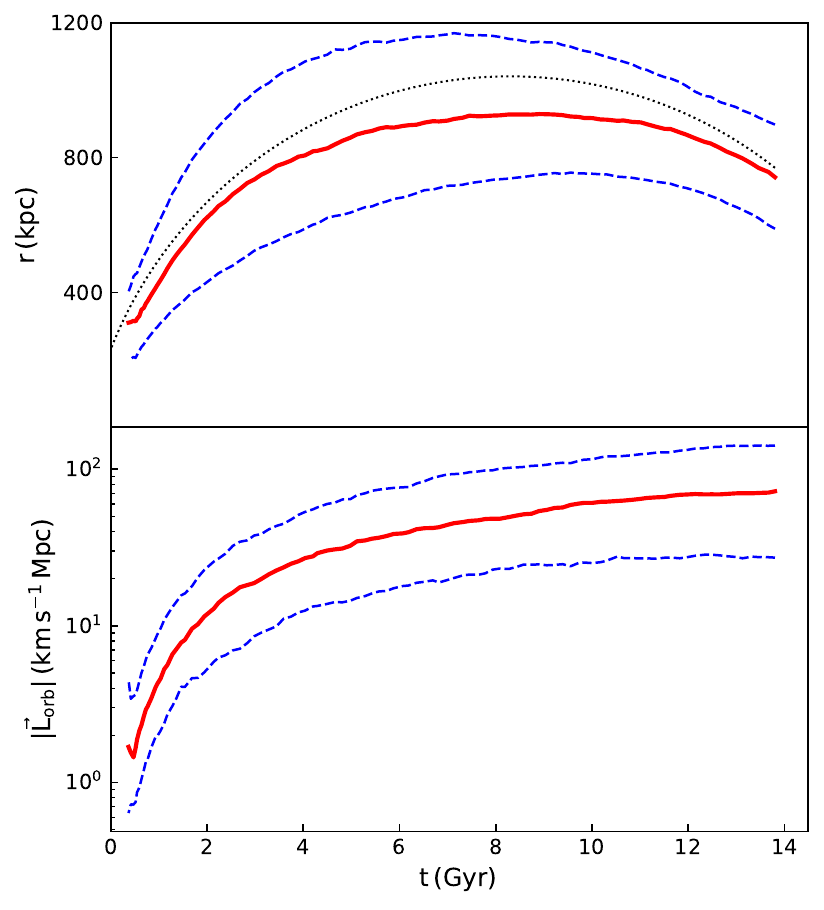}
    \caption{Upper panel shows the cumulative orbit of all 597 pairs where the solid red curve represents the median orbit and the blue dashed curves represent the 68\% c.l. From the cumulative separation between each pair at each snapshot and limiting our perspective to the 68\% c.l. we recover a typical orbit in which the LG is on its first infall. The black dotted curve in the top panel represents the simplified orbit for the LG where $|\vec{L}_{orb}| = 0$, and using the typical present day values of $r_{0} = 770$ kpc, $ \mathrm{V_{rad} = -110\, km \, s^{-1}}$, and $\mathrm{M_{LG} = 3.54 \times 10^{12} \,  {\rm M}_{\odot}}$.  The bottom plot shows the angular momentum growth over time, where we calculated the orbital angular momentum of the pairs at each snapshot. The solid red line shows the median while the dashed blue curves represent the 68\% c.l.}
    \label{fig:cum_orbit_L}
\end{figure}

We calculate the orbital history of the MW-M31 analog pairs identified from the simulation by calculating the relative separation between the MW and M31 at each snapshot, and examining the separation as a function of time and redshift. In order to highlight the diversity of orbital histories, the top panel in Figure~\ref{fig:simorbits} shows orbits for two example LG analogues. The orbital history plotted for the analog pair on the left is an example of an orbit which is perfectly described in the timing model, in which $r \rightarrow 0$ as $t \rightarrow 0$, and the pairs are approaching the pericenter for the first time today. This system has a total mass of $5.3 \times 10^{12} \, \mathrm{M_{\odot}}$, a separation of 970 kpc and the radial and tangential velocities are $-65 \, \mathrm{ km \, s^{-1}}$ and  $37 \, \mathrm{ km \, s^{-1}}$, respectively. For comparison, the right plot shows a pair that has undergone a pericentric passage and is now approaching its pericenter for the second time today. This pair that has undergone a percentric passage has a total mass of $5.6 \times 10^{12} \, \mathrm{M_{\odot}}$, a separation of 651 kpc and the respective radial and tangential velocities are calculated to be $-40 \, \mathrm{ km \, s^{-1}}$ and $144 \, \mathrm{ km \, s^{-1}}$. 

Figure~\ref{fig:simorbits} indicates that the individual LG-analog pairs can have vastly differing orbital histories. To get a sense of the average orbital history for our entire sample, we plot the cumulative distribution of the relative separation for all our pairs in Figure~\ref{fig:cum_orbit_L} (top). This shows that the mean orbit reached an apocenter several Gyr ago, and is approaching the apocenter for the first time. This result was also found in~\citet{2023MNRAS.526L..77S}, and shows that even though an individual orbit may be significantly different, the mean orbit is well-described by the classical timing orbit model. 

To further highlight the range of orbital histories for our entire sample, we note that 77 have undergone a pericentric passage, and are approaching their pericenters for the second time. There are no systems that have had two pericenter passages. Systems that have undergone pericentric passages have a wide range of masses ($\mathrm{\sim  0.8 \times 10^{12} M_{\odot} - 1.0 \times 10^{13} M_{\odot}}$), though they tend to have lower modulus for their radial velocities, $ \gtrapprox -60 \, \mathrm{km \, s^{-1}}$. This, perhaps surprisingly, suggests that it is still possible for a MW/M31 pair to have already had a previous passage. Further, through a visual inspection and anticipating the analysis below, we define ``isolated" pairs as those with a smooth trajectory. We find 15 of 77 pairs with pericentric passages satisfy this smooth criteria, with pericenters ranging from $60 - 700 \mathrm{\, kpc}$.

\subsection{Orbital Angular momentum}
We now move on to consider the orbital angular momentum. We determine the orbital angular momentum from the separation and tangential velocity of the LG-analogs. Figure~\ref{fig:Lkdeonly} shows the probability density of the resulting magnitude of the orbital angular momentum calculated for all 597 pairs (red curve) at $z=0$. We compare this to the blue curve which shows the uncertainty in the angular momentum of the LG calculated from the EDR3 proper motions, and find these two calculations are in good agreement. So, in this sense the observed orbital angular momentum is consistent with our LG-analogue sample.

We then calculate the orbital angular momentum for all snapshots to study how this quantity evolves for LG-like systems. The bottom panel in Figure~\ref{fig:simorbits} shows the example orbital angular momentum growth over cosmic history for the two analogues shown previously. We see both of these pairs, regardless of whether they are on first or second infall, show a steady increase in angular momentum up to the present-day snapshot. To see if this angular momentum growth is a general trend across all LG pairs, we plot the cumulative orbital angular momentum at each snapshot in Figure~\ref{fig:cum_orbit_L} (bottom). It is interesting to note that, when examining the 95\% containment for the orbital angular momentum, the width of this region is fairly constant across time, indicating that this trend is similar for most LG analogues. 

\subsubsection{Origin of $\Vec{L}_{orb}$}
\begin{figure}
    \centering 
	\includegraphics[width = 0.45\textwidth]{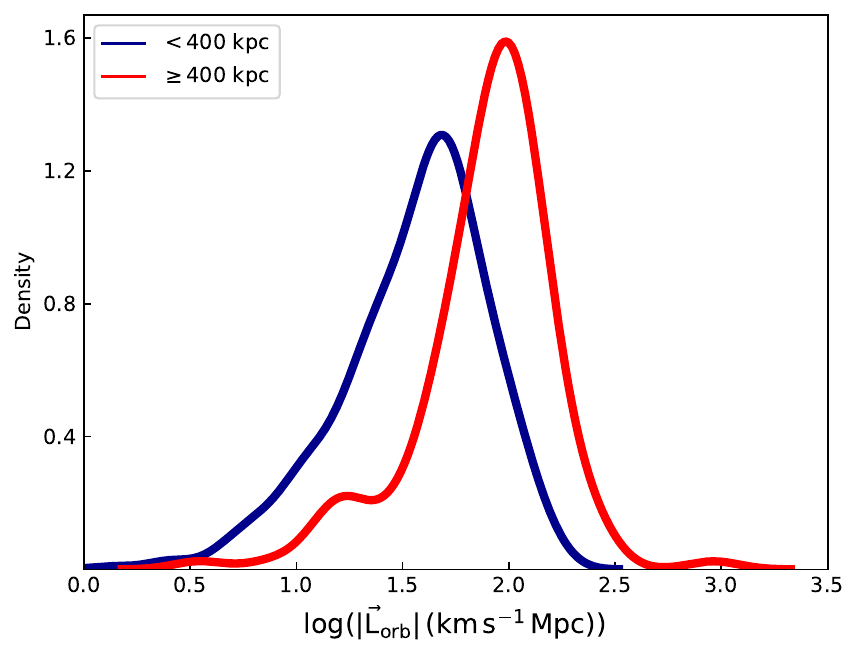}
   \caption{Simulated LG sample split based on its formation distance, with those pairs formed within 400 kpc shown in blue and pairs formed further than 400 kpc in red. We see the further pairs formed, the larger the angular momentum for these systems. The median of $\mathrm{\log_{10}|\vec{L}_{orb} ( \, {\rm km} \, {\rm s}^{-1} \, {\rm Mpc} )| }$ is 1.62 and 1.95 for the $< 400$ kpc and $> 400$ kpc samples, respectively. }
    \label{fig:Lkde_form400}
\end{figure}

\begin{figure*}
\centering 
    \includegraphics[width = 0.49\textwidth]{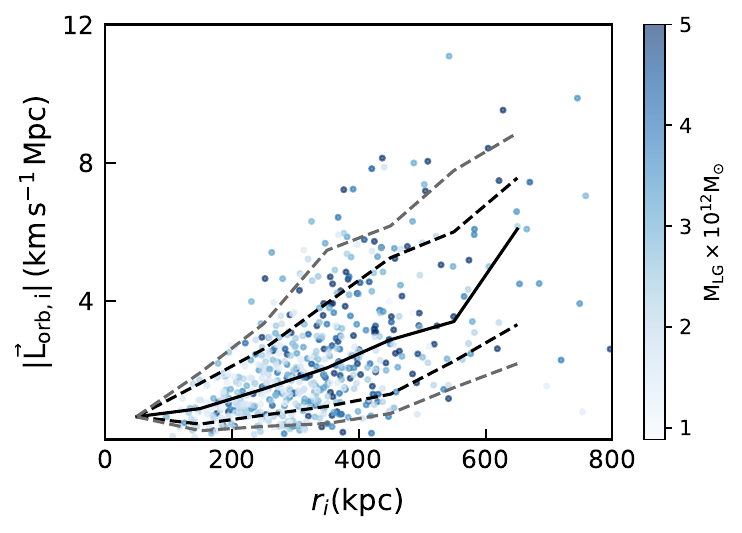}
    \includegraphics[width = 0.49\textwidth]{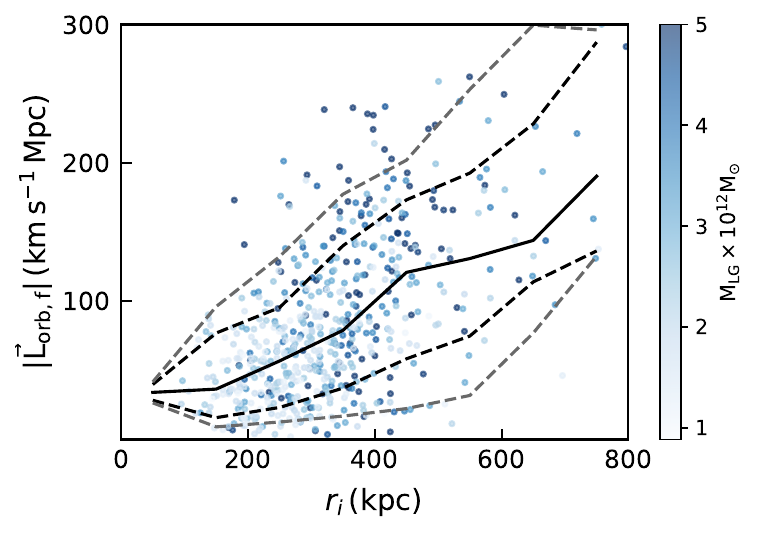}
    \caption{We plot the orbital angular momentum against the formation distance of the pair. The left plot shows the initial orbital angular momentum at the time of formation against the formation distance. The right plot shows the formation distance against the orbital angular momentum calculated at the present day. In both cases we bin by formation distance and plot the running containment intervals, the median shown with the solid curve and the black dashed curve representing the 68\% containment, and the grey dashed curve representing the 95\% containment.  
    }
    \label{fig:LvsRi}
\end{figure*}

\begin{figure}
    \centering 
	\includegraphics[width = 0.50\textwidth]{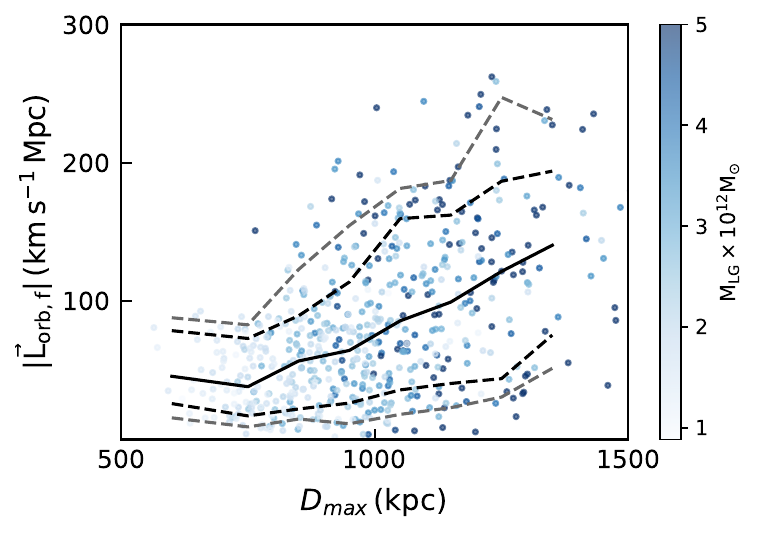}
   \caption{ We plot the orbital angular momentum at $z=0$ vs the maximum distance between the pair. For pairs that follow the canonical timing argument, this distance corresponds to the apocenter. Containment intervals are the same as in Figure~\ref{fig:LvsRi}.}
    \label{fig:Lvsdmax}
\end{figure}

Referring back to Figure ~\ref{fig:LG_orbits}, halos are characterized by a wide range of physical separations, up to as large as $\sim 2400$ kpc. Pairs with a larger magnitude of approach velocity, i.e. more negative values of radial velocity, tend to have larger separations at identification. However, a more significant trend is noted when considering the tangential velocity, which varies with separation at identification in all cases. We examine this trend further with our simulation sample, where we define $r_{i}$ as the separation at the earliest snapshot at which both of the main halos are identified in the system. This can be roughly thought of as the distance the two galaxies were when the LG ``formed" as a system, and is related to the size of the overdensity that collapsed to form the LG~\citep{2016MNRAS.462L..51B}. 
 
 We split the sample of LG-analogs into two, those with physical separations $r_i< 400$ kpc and those with $r_i >400$ kpc, resulting in 135 and 434 pairs, respectively. We found no preference for pairs to be in either of the initial separation bins based on whether they were on their first in-fall or second in-fall. However, Figure~\ref{fig:Lkde_form400} plots the orbital angular momentum of these two initial separation samples, where we see a very notable effect that the sample with larger such formation distances (red curve) have larger orbital angular momentum than those pairs that formed more in closer proximity (blue curve). Thus, the present-day tangential velocity appears to be a reflection of the earliest time in the formation of the LG system. 

To investigate this issue further, we calculate the angular momentum at the earliest snapshot the pair is identified and plot it against the separation at this snapshot. As seen in Figure~\ref{fig:LvsRi} (left), the initial orbital angular momentum increases with larger formation distances. The curves on the plot are constructed by binning in formation distance and computing the median, 68\% and 95\% containment regions shown as the solid black, dashed-black, and dashed-grey lines, respectively. We also examine the orbital angular momentum at $z=0$, and show in Figure~\ref{fig:LvsRi} (right) that this relationship to the initial formation distance is preserved. Note that we investigated the present day orbital angular momentum against the present day separation, and we found no such relationship, which suggests that the tangential velocity drives the relationship shown in Figure~\ref{fig:LvsRi}. 

Finally, we determine the maximum separation of the pair at any point in the simulation, $D_{max}$, and plot this value against the present day orbital angular momentum in Figure~\ref{fig:Lvsdmax}. Figure \ref{fig:Lvsdmax} shows there is also a correlation between the present day orbital angular momentum and the maximum separation. For pairs that follow the canonical timing argument criteria, their maximum separation is the distance of the pair at apocenter: $D_{max}$ = $r_{apo}$. Addditionaly, note that since the MW and M31 are approaching today, their maximum separation must be less than their current separation: $D_{max} < r_{0}$.  According to Figure \ref{fig:Lvsdmax}, apocenters of about 950 kpc correspond to the most likely value of the present day angular momentum $( {\rm in} \, {\rm km} \, {\rm s}^{-1} \, {\rm Mpc} )$ of 64.4 with the respective upper and lower bounds of 113.8 and 26.3. We can divide by the present day separation of 770 kpc to get the median tangential velocity of 83.7 $\mathrm{km \, s^{-1}}$, with 147.8 $\mathrm{km \, s^{-1}}$ and 34.2 $\mathrm{km \, s^{-1}}$as the respective upper and lower bounds. It is interesting to note that the median estimated value of the tangential velocity using this method is $< 2 \mathrm{\, km \, s^{-1}}$ from the most recent measured value. 

\subsection{Orbital Matching}
\label{sec:orb_matching}
\begin{figure*}
\centering 
    \includegraphics[width = 0.44\textwidth]{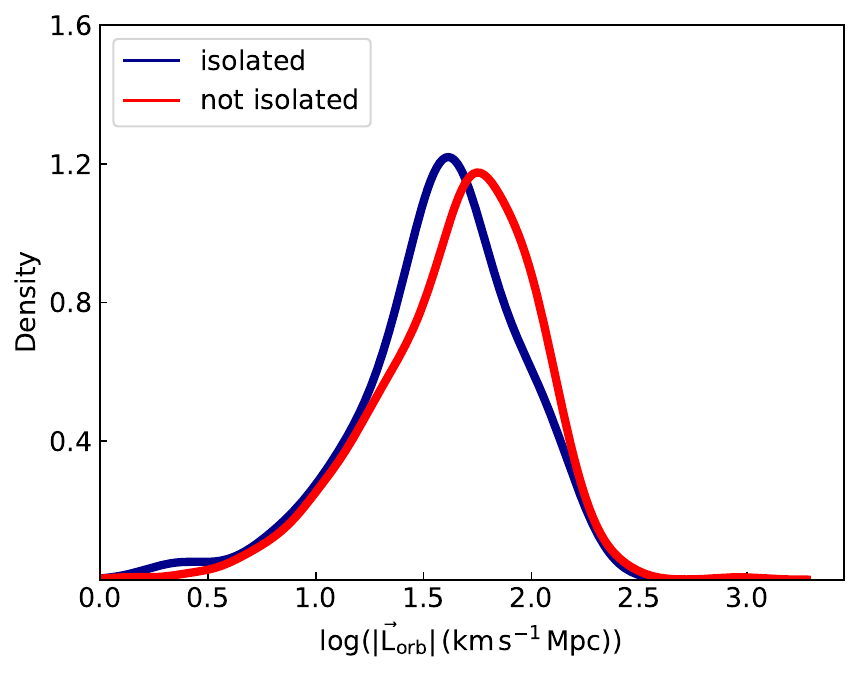}
    \includegraphics[width = 0.51\textwidth]{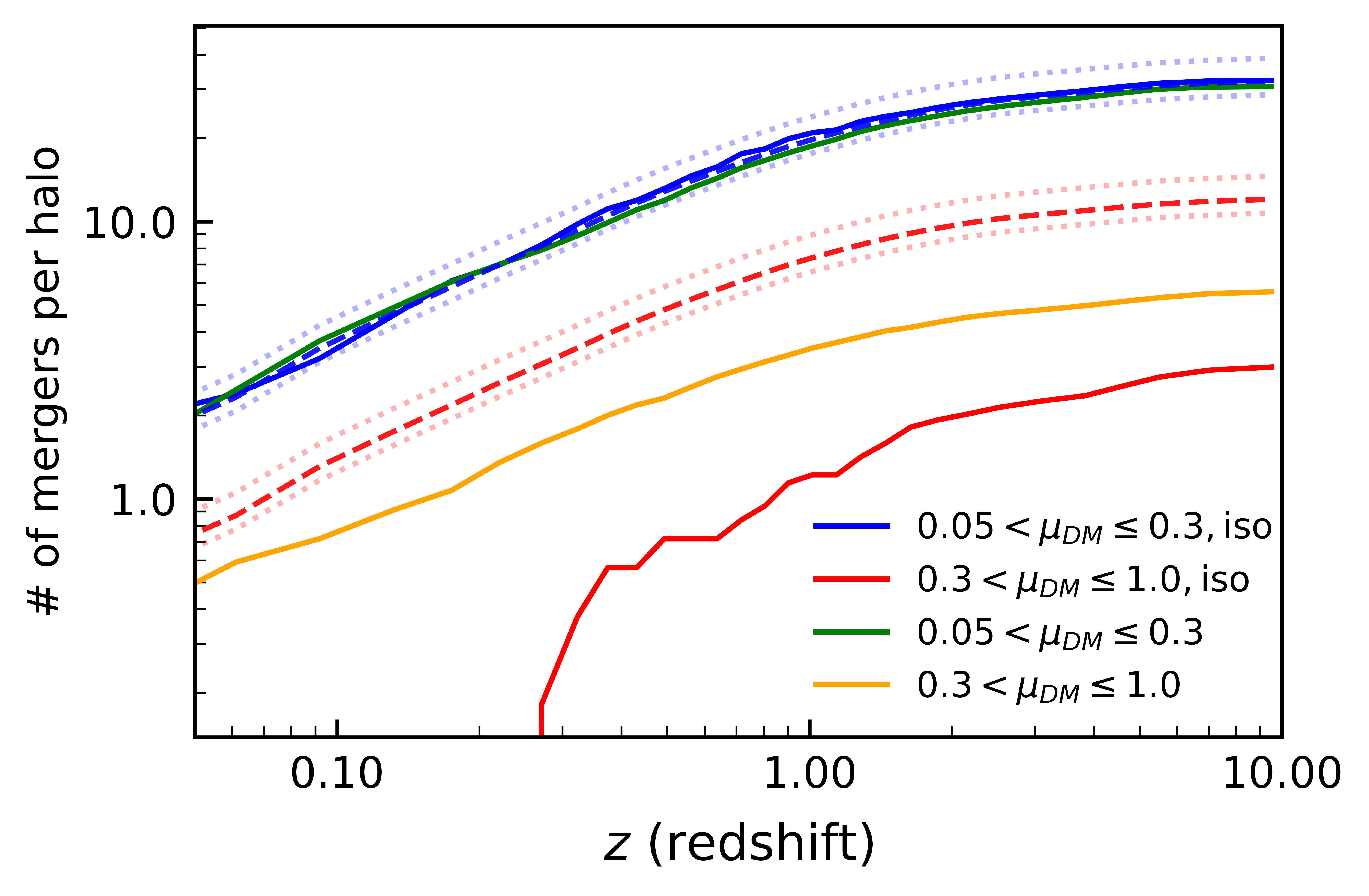}
    \caption{The left plot shows the distribution of the angular momentum for the two samples of Local Group analogues defined by the Keplerian isolation requirement with $C < |0.1|$. The 51 isolated pairs (blue) are shown to have a smaller angular momentum with a median of 1.60 than the 546 pairs which are not isolated by this criteria, which have a median of 1.70. On the right we show the major and minor merger rate for the two samples. The red and blue curves represent the sample isolated by this criteria, and the green and orange curves represent the remainder of the sample. Similar to Figure \ref{fig:all_merger}, the dashed curves represent the merger rate predicted by the fitting function for a given mass range associated with each sample. 
    }
    \label{fig:kep_isolation_strict}
\end{figure*}

With an understanding of typical orbits and angular momenta for our LG analogues, we now move on to examining how well a simple, isolated two-body model describes the orbits. We start from Equation \ref{eq:MTAKEP}, and compare the implied trajectory to the true orbit from the simulation, and thereby determine how well Equation~\ref{eq:MTAKEP} predicts the orbital history of a typical MW/M31 pair. 

Starting from the present day snapshot, we compare the predicted separation from the analytic orbit, $|\vec{r}_a|$, to the true separation in the simulation $|\vec{r}_s| $ by defining the quantity $C = (|\vec{r}_a|- |\vec{r}_s|/)|\vec{r}_s|$. We track each of the orbits from the present day back to when the universe was 3.2 Gyr old ($z=2$). Rather than tracing back to the earliest snapshot, $z\approx20$, we choose 3.2 Gyr in order to remove noise that is introduced due to different halo identification times. To identify matched orbits we consider a strict criteria for $|C| < 0.1$ yielding 51 pairs. If we instead loosen this criteria to $|C| < 0.2$ the sample increases to 169 pairs. However, for this paper we will focus our analysis on the stricter sample of 51 pairs.

We briefly investigate the significance of incorporating the $\Lambda$ term in our orbital predictions. If we instead set $\Lambda = 0$, our analysis reveals only one instance where $|C| < 0.1$. This finding stresses the critical necessity of including the cosmological constant in our models to accurately reconstruct and predict the orbital history of the MW and M31.

Figure \ref{fig:kep_isolation_strict} compares the angular momentum and the merger rate for the sample of 51 pairs, which closely matched the predicted analytical orbit, to the remainder of the sample. In the left panel of Figure \ref{fig:kep_isolation_strict}, we see a slight downward shift in the angular momentum with a median shift from $\mathrm{\log_{10}|\vec{L}_{orb} ( \, {\rm km} \, {\rm s}^{-1} \, {\rm Mpc} )| = 1.70}$ to 1.60 for the 51 systems that match the analytic orbit. As compared to the measured orbital angular momentum derived from the Gaia EDR3 proper motions, the calculated orbital angular momentum of the real LG is consistent with both samples, though the median is in slightly better agreement with that of the non-isolated Local Group pairs.

A more interesting notable result is shown in the right plot in Figure \ref{fig:kep_isolation_strict}, which plots the merger history for each of these samples. Here we show that pairs with $|C| < 0.1$ have had much more quiet recent major merger histories as compared to the sample with $|C| > 0.1$. There appears to be little difference in the minor merger histories for the two samples. This implies that the analytic two-body orbit discussed in Section~\ref{sec:characteristics} is a good proxy for those systems that evolved in isolation, at least in the sense that systems that evolved in isolation are those that have had a quiet major merger history. 

In our analysis, we confirmed that there were no close encounters among the 51 systems that align with the analytic orbit criteria. Out of the 77 pairs identified, only 4 were found to have experienced ambiguous pericentric passages that fell within the subset of 51 systems. Furthermore, all 4 systems were characterized by significant pericentric distances ($ > 700 \, \mathrm{kpc}$). This leads us to conclude that accurately predicting the orbits of systems with very close passages remains a challenge

A question may arise as to whether the MW and M31 are isolated due to the proximity of the massive LMC satellite. Current estimates of the LMC’s mass faction to the MW range from 10\% to 20\% \citep{2019MNRAS.487.2685E,2021ApJ...923..149S,2016MNRAS.456L..54P}, which is considered a minor merger by our definition. The right plot of Figure \ref{fig:kep_isolation_strict}  shows that minor mergers do not affect our ability to predict the orbit; only the recent ($z<0.25$) major ($\mu_{DM} \geq 0.3$) mergers do. By this definition, the LMC is included in this isolation criteria and thus may not significantly affect our ability to predict the orbital history of the MW/M31 using this two-body approximation.

This conclusion remains valid even if the simulation pairs do not represent systems with an LMC-like satellite at present day. This is because our criteria for isolated pairs is to predict the orbit within 10\% accuracy from present day to a universe age of $t=3.2$ Gyr. This traces the orbit far enough back in time to where there must have been an LMC- like object for most of these systems at some point. \citet{Hartl:2021aio} studied these pairs and could not find LMC-like candidates with present-day distances comparable to the observed distance between the LMC and the MW. However despite the simulation sample lacking present-day LMC satellites, due to the time frame covered by our analysis to predict the orbits we can still conclude that the system may still be in isolation with the existence of the LMC, as long as there are no major mergers after $z=0.25$.

\section{Discussion and conclusion}

In recent years, there has been significant progress in the measurements of the relative motion of the MW and M31. Most notably, it is no longer valid to assume that the MW and M31 are on a purely radial orbit, as new proper motions measurements suggest a non-negligible tangential velocity. Furthermore, it has been realized that the LMC may introduce an additional correction to the relative velocity, possibly the nature of the orbit. Accounting for these new measurements, in this paper we study the orbit, angular momentum, and merger rate for MW-M31 like systems, using a combination of an analytic orbit model and simulations of LG-like systems. 
 
We find that the uncertainty on the tangential velocity, combined with the uncertainty in the total mass of the MW and  M31, results in a wide possible range of orbital histories. Additionally, correcting for the LMC influence may alter the approach velocity significantly, which we showed when taken into account, results in a higher probability that the MW and M31 have undergone a pericentric passage and are approaching pericenter for the second time today (See~\citet{2021A&A...656A.129B} for previous discussion of possible past encounters for the MW and M31). 

We use a large sample of $\sim 600$ MW/M31 analog pairs in the IllustrisTNG 300-1 simulation to study their orbital history and orbital angular momentum. We look at all previous simulation snapshots and calculate each pair's orbital history. While we show the average orbit of all pairs implies that the MW and M31 are on first infall, we find 77 (13\%) of these pairs have already undergone a pericentric passage. Out of these 77 pairs, we identified 15 that were isolated by satisfying a smooth criterion, with orbits showing clear pericentric passages with pericenters ranging from 60 kpc to 700 kpc. Although we did not observe a difference in the present-day orbital angular momentum between this sample and the sample of pairs at first infall,  we did find a slight correlation with the merger rate. The pairs that had undergone a pericentric passage generally had a more active merger history.

We use these analog pairs to determine how well the two-body approximation predicts the orbital history between the pair when known kinematic parameters are provided. We find we are only able to predict the orbit within 10\% accuracy for $\sim 10\%$ of the sample, and that we can predict the orbit best for those pairs in which the MW and M31 have not undergone a major merger since $z \approx 0.25$. So, if we can conclude that the MW and M31 have not undergone a merger since $z=0.25$, then we can be more confident in the orbital history we recover using this two-body model. While this is likely true for the merger history of the MW, M31 may have had a more active recent major merger history
 
As a result of our analysis we derive the Local Group orbital angular momentum to be $\mathrm{\log_{10}|\vec{L}_{orb} ( \, {\rm km} \, {\rm s}^{-1} \, {\rm Mpc} )| = 1.80_{-0.21}^{+0.14}}$, and check its consistency with our sample from the $\Lambda CDM$ simulation.  We find that the magnitude of the angular momentum as derived from the methods is in good agreement and that the range of the magnitude of the angular momentum is consistent with what is seen in various simulations which identify LG candidates~\citep{2013ApJ...767L...5F,2016MNRAS.458..900C,Hartl:2021aio}. We then used these pairs to study the evolution of the orbital angular momentum.
We calculate the orbital angular momentum for each pair at each snapshot, and we find that each the orbital angular momentum of the pairs steadily increase up to the present day (Figure \ref{fig:cum_orbit_L} bottom). Though we do not specifically address the origin of this angular momentum, one possibility is tidal torques acting to the LG during the course of its evolution~\citep{1984ApJ...286...38W,2002MNRAS.332..325P}.  

We found a wide range of initial separations between the pair at the snapshot of their first identification in the simulation. We split our sample based on those with initial distance at the time of identification as $> 400$ kpc and those with $< 400$ kpc. While we did not find a correlation between the present day orbital angular momentum and the present day separation, we did interestingly find a correlation such that pairs with a larger initial separation have a larger orbital angular momentum today. The fact this correlation is preserved through cosmic time strongly suggests that the magnitude of the orbital angular momentum is an imprint of the conditions on the system at the time of formation.

The finding that pairs that evolved from larger separations have a larger orbital angular momentum may also reflect the influence of tidal torques on these systems. Most simply, the magnitude of the tidal torque is proportional to the quadrupole moment of the proto-LG mass distribution, in addition to the quadrupole moment of the external gravitational potential. The fact that all systems indicate a steady increase in orbital angular momentum suggests that a build of tidal torques occurs for all systems. But the fact the initial and final angular momentum are both correlated to the formation distance, as seen in Figure \ref{fig:LvsRi}, indicates the magnitude may be set by the initial formation conditions of the LG.

Our results have highlighted that the uncertainty in tangential velocity measurement and the total mass of the MW and M31 preclude, at this stage, definitive answers to the questions of the LG formation history. The kinematics and mass measurements may improve with future measurements from Gaia, which are expected to improve in precision due to single-epoch astrometry. Further, improved measurements of the LG kinematics would provide information on the direction of the LG orbital angular momentum vector, which may be compared to the matter distribution in the local volume. These topics will be addressed in a forthcoming study. 

\section*{Acknowledgements}
We are grateful to Jorge Penarrubia and Carton Zeng for discussions on this paper. We acknowledge support from DOE Grant de-sc0010813, and by the Texas A\&M University System National Laboratories Office and Los Alamos National Laboratory. 


\section*{Data Availability}
The data underlying this article will be shared on reasonable request to the corresponding author.



\bibliographystyle{mnras}
\bibliography{example} 



\bsp	
\label{lastpage}
\end{document}